\documentclass[journal]{IEEEtran}
\usepackage[dvipsnames]{xcolor}
\usepackage{amsmath,amsfonts}
\usepackage[switch]{lineno}
\usepackage{algorithm2e}
\usepackage{subcaption}
\usepackage{textcomp}
\usepackage{stfloats}
\usepackage{verbatim}
\usepackage{graphicx}
\usepackage{multicol}
\usepackage{multirow}
\usepackage{hyperref}
\usepackage{booktabs}
\usepackage{tabularx}
\usepackage{balance}
\usepackage{siunitx}
\usepackage{amsmath}
\usepackage{amssymb}
\usepackage{lipsum}
\usepackage{braket} 
\usepackage{array}
\usepackage{float}
\usepackage{cite}
\usepackage{url}
\usepackage{mwe}

\SetKwComment{Comment}{/* }{ */}
\RestyleAlgo{ruled}
\begin{document}

\title{SEN12-WATER: A New Dataset for Hydrological Applications and its Benchmarking}

%SEN1-2DWATER: Addressing Climate Change and Water Scarcity with a Comprehensive Dataset
%SEN1-2DWATER: A New Dataset to Tackle Climate Change and Water Scarcity Issues
%SEN1-2DWATER: Advancing Drought Analysis with Multimodal Data

\author{\IEEEauthorblockN{Luigi Russo, \IEEEmembership{Student Member, IEEE},
Francesco Mauro, \IEEEmembership{Student Member, IEEE}, Alessandro Sebastianelli, \IEEEmembership{Member, IEEE},  
Paolo Gamba, \IEEEmembership{Fellow, IEEE}, Silvia L. Ullo, \IEEEmembership{Senior Member, IEEE} 
}
\thanks{Luigi Russo and Paolo Gamba are with the Department of Electrical, Computer and Biomedical Engineering, University of Pavia, 27100 Pavia, Italy (e-mail: \{luigi.russo02, paolo.gamba\}@universitadipavia.it)\\
Francesco Mauro and Silvia Liberata Ullo are with the Department of Engineering, University of Sannio, 82100 Benevento, Italy (email: \{f.mauro@studenti., ullo\}@unisannio.it).\\ 
Alessandro Sebastianelli is with the European Space Agency, Frascati, Italy (email: Alessandro.Sebastianelli@esa.int)}}

\date{}
\maketitle
\markboth{Submitted to IEEE Transactions on Geoscience and Remote Sensing}{}

\begin{abstract}
Climate change and increasing droughts pose significant challenges to water resource management around the world. These problems lead to severe water shortages that threaten ecosystems, agriculture, and human communities. To advance the fight against these challenges, we present a new dataset, SEN12-WATER, along with a benchmark using a novel end-to-end Deep Learning (DL) framework for proactive drought-related analysis. The dataset, identified as a spatiotemporal datacube, integrates SAR polarization, elevation, slope, and multispectral optical bands. Our DL framework enables the analysis and estimation of water losses over time in reservoirs of interest, revealing significant insights into water dynamics for drought analysis by examining temporal changes in physical quantities such as water volume. Our methodology takes advantage of the multitemporal and multimodal characteristics of the proposed dataset, enabling robust generalization and advancing understanding of drought, contributing to climate change resilience and sustainable water resource management. The proposed framework involves, among the several components, speckle noise removal from SAR data, a water body segmentation through a U-Net architecture, the time series analysis, and the predictive capability of a Time-Distributed-Convolutional Neural Network (TD-CNN). Results are validated through ground truth data acquired on-ground via dedicated sensors and (tailored) metrics, such as Precision, Recall, Intersection over Union, Mean Squared Error, Structural Similarity Index Measure and Peak Signal-to-Noise Ratio.
\textcolor{red}{Data and code will be opened upon acceptance of the paper. Limited access can be granted during the review process upon request.}   
\end{abstract}


\section{Introduction}
\IEEEPARstart{C}{limate} changes are having an impact on the occurrence of extreme events, such as droughts and water scarcity, on one hand, and floods and landslides on the other. It is worth highlighting that drought and floods are two events with different dynamics developing on different timescales. In this work, it has been chosen to focus on drought conditions to provide clear evidence of climate change and its impact on the reduction of water bodies and the water scarcity.

Water availability is becoming less predictable in many places in the world and, in some of them, droughts are exacerbating water scarcity, thus negatively impacting people's health and productivity and threatening sustainable development and biodiversity. In this context, it is crucial to locate water bodies and reservoirs, to observe their changes over time, to detect hydrological patterns (patterns and behaviors of water resources in a specific area) and their impact on changes in soil and water content, in order to monitor and predict the occurrence of such catastrophic events \cite{yang2020general}. The first fundamental step in building a DL framework for accurately identifying water basins and performing detailed analysis, is having a suitable dataset available \cite{linke2019global, sebastianelli2021automatic}. To this end, satellite data and their processing can offer support and clear evidence of the changes occurring on Earth, providing the overall picture and gathering long-term data series to comprehend the consequences of these changes \cite{teodoro2022role}. 

For these purposes, this article proposes a new multisource and multitemporal dataset that is unique in several aspects compared to other already available datasets, as discussed ahead. Moreover, in combination with the dataset, a novel end-to-end framework is presented, including as main internal blocks two different neural networks: one U-Net architecture for assessing water segmentation, and a Time Distributed (TD)- Convolutional Neural Network (CNN) fed with the output of the U-Net and able to accomplish the next frame prediction.

The objective is to offer a complete tool to policymakers and research community, with a specific dataset and a suitable framework together, able to allow estimation of water losses for reservoirs over a certain period. Thus, this study introduces several key contributions and novel aspects:

\begin{itemize}
    \item \textbf{SEN12-WATER Dataset:} introduction of a novel multisource and multitemporal dataset, SEN12-WATER, which combines SAR polarization, elevation, slope, and multispectral optical bands. This dataset provides a comprehensive spatiotemporal datacube for analyzing water bodies, which is unique compared to existing datasets.
    \item \textbf{End-to-End Deep Learning Framework:} A new end-to-end Deep Learning (DL) framework is developed for the analysis and estimation of water losses over time. This framework includes advanced models such as a ResNet for speckle noise removal from Sentinel-1 (S1) images, a U-Net for water body segmentation, and a Time Distributed-Convolutional Neural Network (TD-CNN) for predicting future water masks.
    \item \textbf{Validation with Ground Truth Data:} The results of the proposed framework are validated using appropriate Ground Truth data acquired on-ground via dedicated sensors or stations and tailored metrics, demonstrating the practical applicability and effectiveness of the methodology in real-world scenarios.
    \item \textbf{Contribution to Climate Change Resilience:} By providing detailed insights into water dynamics and enabling proactive measures against drought-related issues, the study contributes to climate change resilience and sustainable water resource management. The framework helps in understanding the impacts of climate change on water bodies and supports informed decision-making for mitigating water scarcity.
\end{itemize}

These contributions collectively advance the state-of-the-art in water drought, offering new tools and methodologies that address current limitations and open up new possibilities for research and application in environmental science, while also providing valuable support for policymakers. This enables better informed decision making and more effective strategies to address environmental challenges.

The article is organized as follows. The next section introduces and discusses related works found in the state of the art (SOTA). Section III introduces the new dataset in detail, explaining its peculiarities. In Section IV the proposed framework is described with the different networks used to achieve the needed steps, demonstrating how it works as an end-to-end network able to produce valuable insights as planned. In Section V the experiments carried on to demonstrate the performance of what proposed are introduced and discussion developed. Conclusions in Section VI end the work.  

\subsection{Related works}\label{related_works_section}

\begin{table*}[!ht]
    \centering
    \caption{Comparison with other SOTA datasets. Our dateset is the only one that contains S1, S2, and SRTM at $10\unit{\meter}$. Regarding the number of samples it is placed among the medium-size dataset, making it more portable, even if its modularity allows straightforward expansion.}
    \label{dataset_comparison_sota_table}
    \resizebox{1.9\columnwidth}{!}{
        \begin{tabular}{llcccrr} 
            \toprule
            \textbf{Paper/Dataset} & \textbf{Satellite} & \textbf{Multisensor} & \textbf{Multitemporal} & \textbf{Resolution} & \textbf{Size} & \textbf{Patch Size}\\
            \midrule            
            Sui et al.\cite{Sui2022} & S2 & X & X & 10m & \num{54304} & - \\
            Zhao et al. \cite{zhao2024urbansarfloods} & S1 & X & $\surd$ & 10m & \num{8879}  & $512px\times512px$ \\
            Tottrup et al. \cite{tottrup2022surface} & S1, S2, Landsat-8 & $\surd$ & $\surd$ & 10m & 7980 & -\\
            Wieland et al. \cite{wieland2023s1s2} & S1, S2, SRTM & $\surd$ & $\surd$ & 10m & $\sim$\num{100000} & $256px\times256px$\\
            Feng et al. \cite{Feng2016} & Landsat7 ETM+ & X & X & 30m & \num{8756} & -\\
            Pekel et al. \cite{Pekel2016} & Landsat & X & $\surd$ & 30m & \num{3000000} & - \\
            \midrule
            \textbf{SEN12-WATER} & \textbf{S1, S2, SRTM} & $\boldsymbol{\surd}$ & $\boldsymbol{\surd}$ & \textbf{10m} & $\boldsymbol{\sim}$ \textbf{\num{205300}} & $\mathbf{64px\times 64px}$\\
            \bottomrule
        \end{tabular}
    }
\end{table*}

As highlighted in the previous section, droughts as a consequence of climate change are exacerbating water scarcity, impacting people's lives and threatening sustainable development. It has become crucial to predict water availability, losses and dynamics of reservoirs, in general. In this Section, SOTA analysis is carried on to verify which specific datasets have been made available, together with related algorithms for processing them. The description of what is proposed comes in the end, highlighting the differences, the novelties and suitability for the aimed objectives with respect to what by now available. 

From SOTA analysis, several datasets have been identified. A crucial requirement to assess climate change's impact on water basins is to have available temporal data sequences with suitable characteristics. Therefore, this aspect is carefully researched and verified. Table \ref{dataset_comparison_sota_table} shows the main features of these datasets used in related works to map the water presence through the employment of satellite images. 

In particular Pekel et al. \cite{Pekel2016} propose a  three million Landsat satellite images spanning $32$ years, in order to quantify changes in global surface water at $30\unit{\meter}$ resolution. However, this work relies only on optical data, which on the one hand can enhance the evaluation with the detailed information contained within each spectral band but, on the other hand, can be compromised in case of adverse climatic conditions, such as cloud coverage, which can easily occur within dense time series that follow the evolution of a reservoir over multiple years. Zhao et al. \cite{zhao2024urbansarfloods} introduce the UrbanSARFloods dataset, designed for flood mapping in both urban and open areas using S1 SLC-based SAR data. The dataset includes pre- and post-event SAR intensity and interferometric coherence imagery. It consists of $\num{8879}$ $512\times 512$ pixel chips covering $\num{807500} \unit{\kilo\meter}^2$ from $18$ flood events across various continents.

The creation of the dataset involves both semi-automatic and manual labelling techniques. Semi-automatic labelling employed conventional remote sensing methods, including hierarchical split-based change detection for open flooded areas and thresholding on differential interferometric coherence for urban flood regions. Manual labelling was performed using high-resolution optical data to ensure precise annotation.

Furthermore, regarding the applications of the dataset, UrbanSARFloods is intended to serve as a benchmark for evaluating state-of-the-art convolutional neural networks in the segmentation of flood areas. The dataset also highlights the challenges associated with urban flood detection, particularly due to data imbalance and the constraints of a relatively small dataset.

A combination of SAR and optical data is proposed by Wieland et al.  \cite{wieland2023s1s2}. They introduced S1S2-Water, which is a global reference dataset for semantic segmentation of surface water bodies based on S1 and Sentinel-2 (S2) satellite images. The dataset consists of $65$ pairs of S1 and S2 images with quality-checked binary water masks. Samples are drawn globally based on the S2 tile-grid ($100 \unit{\kilo\meter} \times 100 \unit{\kilo\meter}$), considering predominant land cover and the availability of water bodies. Each sample is complemented with metadata and a digital elevation model (DEM) raster from the Copernicus DEM. The images were acquired between May 21, 2018, and November 26, 2020, covering almost all months of the year, with no samples available for December. In this work, all experiments are carried out using a U-Net architecture, applied separately to Sentinel-1 and Sentinel-2 data without any data fusion. Sentinel-2 images serve as the initial dataset for annotating water masks. The annotation process is semi-automated, where water bodies are first identified using a threshold procedure based on the Normalized Difference Water Index (NDWI).

This is another critical aspect, as the use of classical threshold-based indices such as NDWI can help in determining the presence of water but, as outlined in \cite{ndwi_critical_issue}, the arbitrariness with which this threshold is set can generate different results, depending on the proportions of subpixel water/non-water components. In a similar way, other works develop a semi-automatic procedures based threshold and classical indices \cite{Feng2016} or assisted by visual interpretation \cite{Sui2022}. Furthermore, Tottrup et al. discuss the creation and application of a dataset for surface water mapping using satellite observations from Sentinel-1, Sentinel-2, and Landsat-8 \cite{tottrup2022surface}. The dataset includes images acquired from July 2018 to June 2020 over diverse test sites with varying topographies and water bodies. The study utilized ancillary datasets such as Digital Elevation Models (DEMs) and pre-existing surface water maps, ensuring they were publicly available. The application involved implementing workflows in Google Earth Engine to facilitate transferability and reproducibility. One approach combined histogram-thresholding and edge-detection methods to estimate monthly surface water extent from cloud-free satellite scenes, creating binary land and water maps for each scene. This method used classical indices like the Normalized Difference Water Index (NDWI) for optical scenes and specific bands  combination for SAR scene.

In recent years, advancements in remote sensing technologies have revolutionized waterbody detection and monitoring. Xiang et al. \cite{10057261} introduce Dense Pyramid Pooling Module (DensePPM) for enhancing water body identification in aerial images through semantic segmentation networks, while Xia et al. \cite{xia2021dau} propose a dense skip connections network with multi-scale features fusion and attention mechanism to effectively identify water areas amidst complex backgrounds. Hertel et al. \cite{hertel2023probabilistic} compare variational inference-based Bayesian CNN (BCNN) with Monte Carlo dropout network (MCDN) for deriving water extent estimates, emphasizing the importance of reliable uncertainty quantification. Furthermore, Gonzàlez et al. \cite{gonzalez2024validation} introduce an innovative approach centered around a Match-up Database (MDB) to simplify validation analysis of satellite-derived water products, enhancing data validation efficiency. Ma et al. \cite{ma2023local} develop a local feature search network (DFSNet) with discarding attention module (DAM) for accurate water body segmentation, showcasing significant improvements in segmentation accuracy. Hou et al. \cite{hou2024glolakes} expand spatio-temporal dynamics measurement of water storage in lakes and reservoirs using high-resolution satellite and altimetry data, enabling global freshwater assessment. Yuan et al. \cite{yuan2024satellite} focus on monitoring water levels in the Qingcaosha estuarine reservoir using Landsat-8 and S2 images, demonstrating high accuracy and reliability in satellite-derived results. Li et al. \cite{li2024glh} introduce the GLH-water dataset for global surface water detection, showcasing its strong generalization and practical utility. Lastly, Zhong et al. \cite{9852482} propose NT-Net, an end-to-end semantic segmentation network, for automatic lake water extraction from remote sensing images, addressing challenges and improving extraction coherence and comprehensiveness. These studies collectively contribute to advancing water resource management and monitoring on a global scale.

\section{Data}
Our multimodal and multitemporal dataset, built upon previous work \cite{mauro2023sen2dwater}, captures all complementary characteristics necessary for mapping water bodies effectively \cite{kulkarni2020pixel}, further enhanced by incorporating information on slope and elevation \cite{wieland2023s1s2}. Our dataset follows the \textit{DiRS properties} (diversity, richness, and scalability), which are introduced in the recent guideline for building remote sensing benchmark datasets by Long et al. \cite{long2021creating}. In Table \ref{tab:dirs_properties}, the definition of each property is outlined together with the motivation behind why the aforementioned dataset fulfils these criteria.

\begin{table}[!ht]
    \centering
    \caption{Definitions and motivations for DiRS properties \cite{long2021creating} in the dataset.}
    \label{tab:dirs_properties}
    \resizebox{1\columnwidth}{!}{
        \begin{tabularx}{\textwidth}{lX}
            \toprule
            \textbf{\huge Property} & \textbf{\huge Our Dataset} \\~\\
            \midrule
            \textbf{\huge Diversity} 
            & \huge The dataset is diverse because it encompasses a wide range of spatial and temporal characteristics, which capture variations in atmospheric conditions, landscapes, and water body features. Slope and elevation models are also included to provide additional environmental variability, ensuring comprehensive representation. \\~\\
            \midrule
            \textbf{\huge Richness} 
            & \huge The dataset is rich in features, including radar backscatter data from S1 and surface reflectance data from S2, providing diverse spectral information across 13 different bands. This allows for the detailed analysis of radiometry and spatial resolution. Moreover, slope and elevation data contribute to understanding water flow, drainage patterns, watershed delineation, and water accumulation areas. \\~\\
            \midrule
            \textbf{\huge Scalability} 
            & \huge The dataset is designed for scalability, allowing for easy expansion and updates. Data from July 2016 to December 2022 were collected using an automated Google Earth Engine script, with provisions for continual additions. The data formats and storage structures support the seamless integration of new annotations and images. \\
            \bottomrule
        \end{tabularx}
    }
\end{table}

\begin{figure*}[!ht]
    \centering
    \includegraphics[width=2\columnwidth]{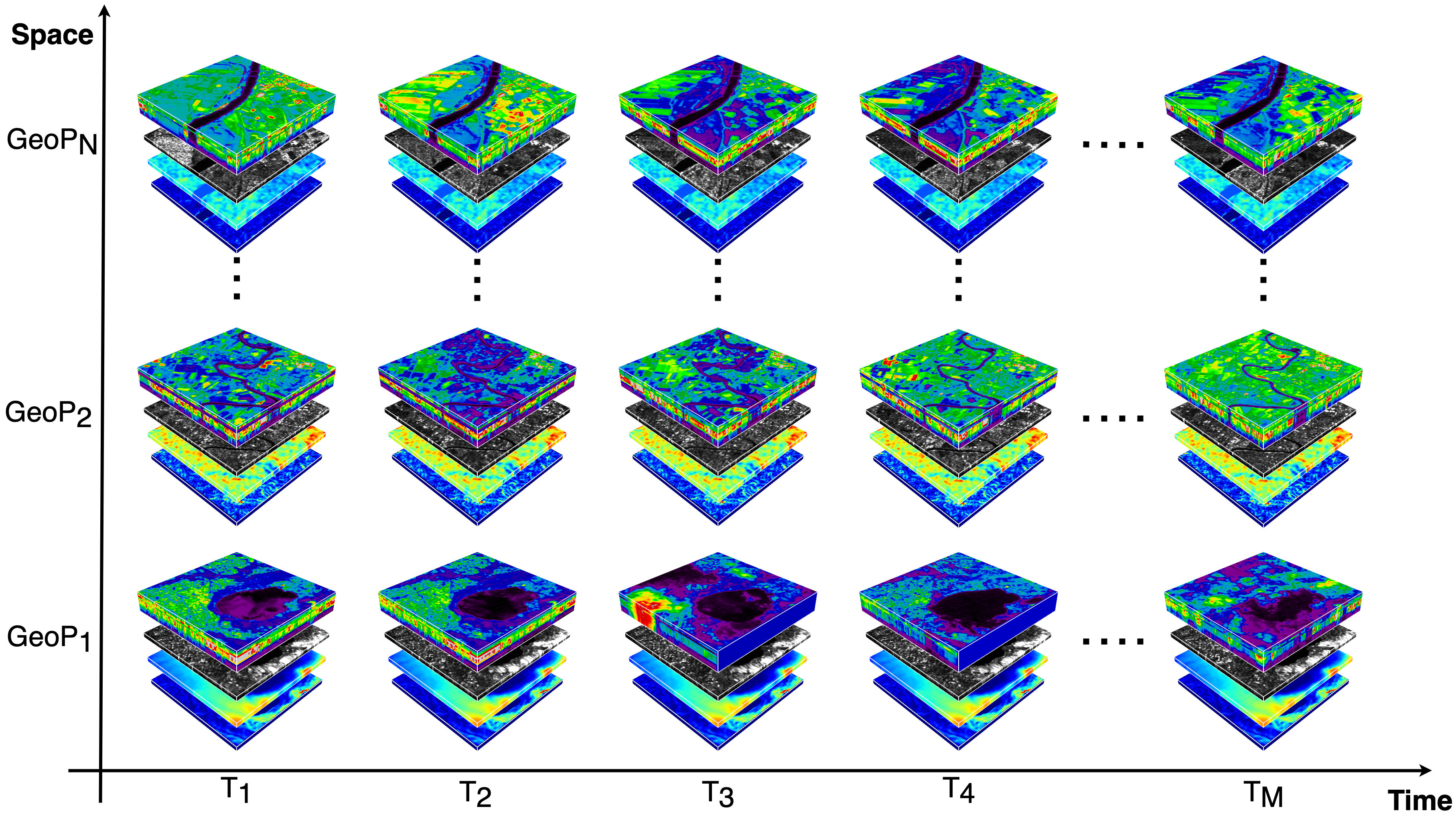}
    \caption{Visualization of the SEN12-WATER Dataset. On the y-axis are different geographical locations, and on the x-axis are the time series images for each location. Each cube plot shows the compound product of the S1 polarizations (VV+VH) and S2 spectral bands (R+G+B+NIR). The datacube is finally enriched with slope and elevation information for each geolocation. This plot has been produced using \href{https://alessandrosebastianelli.github.io/opensv/pyosv.html}{python opensv}.}
    \label{fig:dataset_timeseries}
\end{figure*} 

What sets this dataset apart is its innovative fusion of S1 (\href{https://sentinels.copernicus.eu/web/sentinel/user-guides/sentinel-1-sar/revisit-and-coverage#:~:text=A%20single%20SENTINEL%2D1%20satellite,repeat%20cycle%20at%20the%20equator.}{ESA: S1}), S2 (\href{https://sentinel.esa.int/web/sentinel/missions/sentinel-2}{ESA: Sentinel-2}), slope, and elevation data from Shuttle Radar Topography Mission (SRTM, \cite{farr2007shuttle}). These important characteristics and information will represent a valuable step forward in the following training of the introduced models. The presence of S1 and S2 data enables comprehensive analysis in all atmospheric conditions, with S1 SAR data, and detailed insights with S2 high-resolution multispectral data. Furthermore, the additional information on slope and elevation will further improve the ability to identify the water basins for each area of interest \cite{wieland2023s1s2}.

For our first version of the dataset, we decided to collect data from July 2016 up to December 2020 over Italy and Spain (\textcolor{red}{please refer to project \href{https://unisannioeolab.github.io/SEN12-WATER/}{GitHub} page for geographical distribution}), with a sample every two months. This decision was mainly driven by the fact that during this period it is possible to observe severe anomalies in terms of drought recorded in these regions. The result is a series of $39$ acquisitions for each of $329$ unique geographical extents (plus $329$ layers of slope and elevation) defined by latitude and longitude coordinates, where each of these areas covers a spatial expanse of  $3 \unit{\kilo\meter} \times 3 \unit{\kilo\meter}$. Each sample of the dataset can be defined as $X_{lat, lon} \in R^{T \times W \times H \times B}$ where \textit{lat} and \textit{lon} are the center coordinates, \textit{T} the length of the temporal sequence, \textit{W} the width of the image, \textit{H} the height of the image, \textit{B} the number of bands of the image. \textit{T} is equal to $39$, considering a sample each two months for timeline of almost six years, \textit{W} and \textit{H} are equal to $300\text{ }px$ considering that the images were acquired over a polygonal area of $3 \unit{\kilo\meter} \times 3 \unit{\kilo\meter}$, with a spatial resolution of 10m and $B = 8$ (\textit{VV+VH+R+G+B+NIR+SLOPE+ELEVATION}). The built dataset is depicted in Figure \ref{fig:dataset_timeseries}, with $N$ denoting the count of distinct geographical points $(GeoP_n)$ and $M$ representing the length of the time series. To obtain it, an automated script was devised utilizing the \textit{Google Earth Engine} (GEE) platform to retrieve S1, S2, Slope and Elevation data for selected water basins. Furthermore, the script automatically identified and selected the least cloudy samples within this time-frame as part of the dataset. Each image $GeoP_n^{(m)}$ comprises all $13$ bands of Sentinel-2, with bands possessing spatial resolutions of $20 \unit{\meter}$ and $60 \unit{\meter}$ being resampled to $10 \unit{\meter}$ and the two polarizations, VV and VH, of S1 with $10 \unit{\meter}$ of spatial-resolution. It is noteworthy to mention that starting from January 25, 2022, S2 scenes with a PROCESSING\_BASELINE of $04.00$ or higher underwent a DN range adjustment of $1000$ (\href{https://developers.google.com/earth-engine/datasets/catalog/COPERNICUS_S2_HARMONIZED}{Harmonized Sentinel-2 MSI: MultiSpectral Instrument, Level-1C}). Consequently, we transitioned to the HARMONIZED collection within GEE, as it accommodates this range shift.

\section{Methods}

\begin{figure*}[!ht]
    \centering
    \includegraphics[width=2\columnwidth]{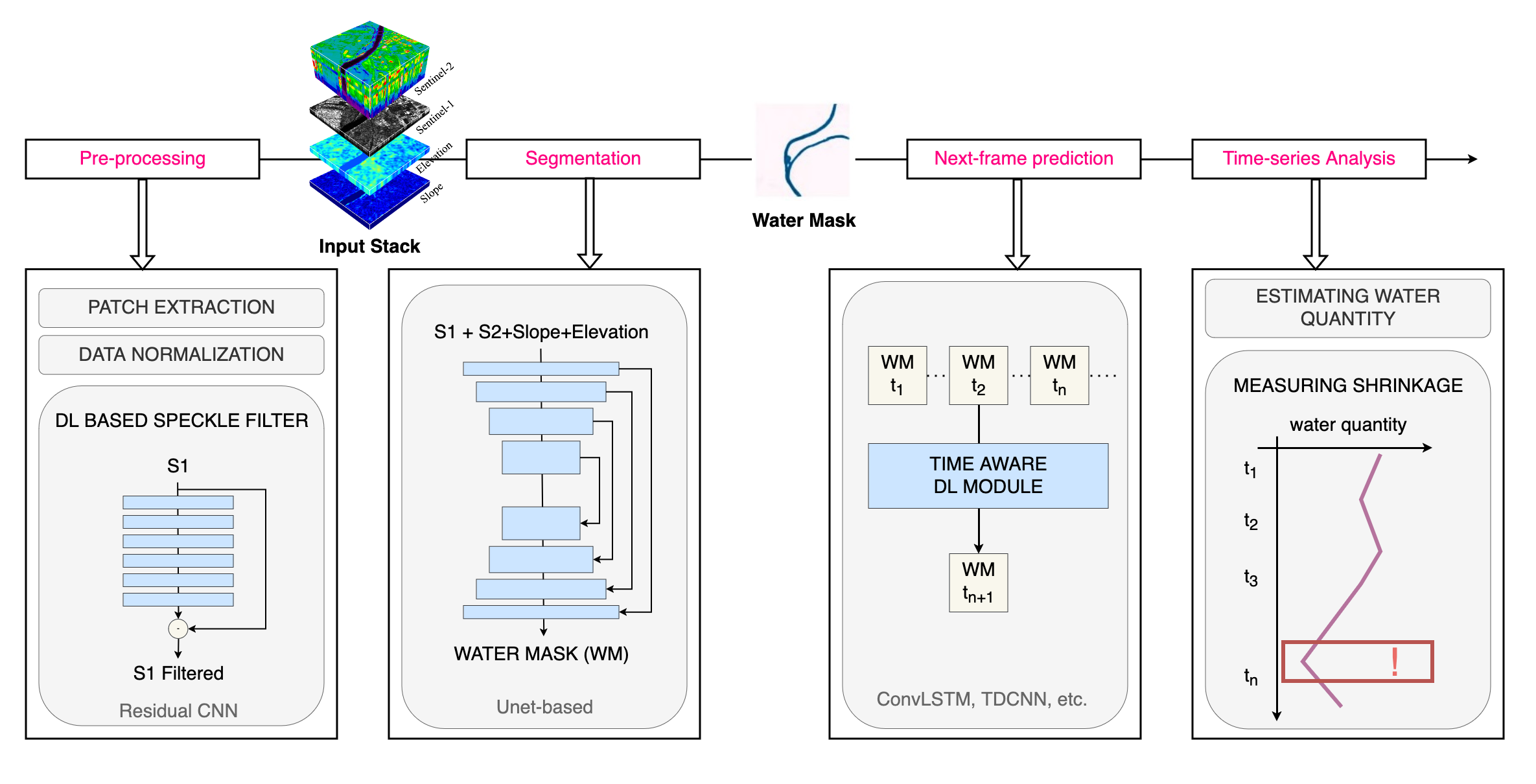}
    \caption{Proposed end-to-end framework and methodology composed of several blocks: segmentation of water basins, prediction of future water and drought masks and evaluation of the quantity of water pixels within each segmented mask.}
    \label{fig:method}
\end{figure*}

The proposed approach rotates around the analysis of multi-source temporal sequences derived from our SEN12-WATER dataset. Our end-to-end model firstly create a segmentation mask, to detect and outline water content from the given image, next it forecast the next frame, and finally it measure the fluctuations in water content, allowing proactive measures against potential drought-related issues.

To effectively perform these tasks, the first step is to accurately identify and track water basins over time, utilizing the comprehensive data collection previously discussed. Before segmenting the water bodies, it was necessary to complete several preprocessing steps, particularly focusing on reducing speckle noise in the SAR data. Once the data was preprocessed, segmentation was applied to the entire dataset to identify the water basins. As an improvement to the approach presented in \cite{igarss2023}, a fully AI-driven method for water volume analysis and forecasting was introduced. In this approach, water masks for each time step are generated using the weights from the top-performing model described in section \ref{unet_section}. Unlike the previous method, which relied on traditional theoretical indices like NDWI to calculate water masks, the new method generates probability maps using a Deep Learning model. As a result, the inputs used for testing these three next-frame prediction frameworks are of higher quality, leading to improved overall outcomes.

Components of our end-to-end method are detailed in the following sections.

\subsection{Pre-processing}
As part of our pre-processing component, despeckling techniques play a crucial role in the accurate identification of water bodies \cite{feng2022mapping}. The use of despeckling methodologies significantly improves the quality and usability of the dataset for mapping and monitoring water basins. The DL methodology for speckle filtering derives from the one presented by Sebastianelli et al. \cite{sebastianelli2022speckle}, appositely adapted to cope with our dataset. It consists of a ResNet model architecture where speckle is subtracted from the input image through a skip connection and a subtraction layer. This choice was driven mainly because: $1)$ data and code are open-access and well details (including tutorials and $2)$ the method is specifically designed to work with the same S1-GRD data used in our study, thus allowing to keep SAR statistical proprieties unchanged \cite{sebastianelli2022speckle}, achieved because of the residual nature and of the following loss function:

\begin{equation}
    \centering
    \begin{split}
        \mathcal{L}_{SPE} &= \mathcal{L}_{MSE} + \mathcal{L}_{SSIM} +\mathcal{L}_{TV}\\&=\alpha_1 \frac{1}{N}\sum_{i=1}^{N}\left[y_{i} - \hat{y}_i\right]^2\\ &+ \beta_1 \frac{(2\mu_y\mu_{\hat{y}} + c_1)(2\sigma_{y\hat{y}}+c_2)}{(\mu^2_y+\mu^2_{\hat{y}}+c_1)(\sigma^2_y+\sigma^2_{\hat{y}}+c_2)}\\ &+ \gamma_1\sum_{w=1}^{W}\sum_{h=1}^{H}\sqrt{(y_{w+1,h}-y_{w,h})^2+(y_{w,h+1}-y_{w,h})^2}
    \end{split}
\end{equation}

where $y$ is the predicted speckle filtered and $\hat{y}$ is the ground truth image respectively; this loss includes the Euclidean distance, the SSIM\footnote{Structural Similarity Index Measure: this represent the second term of the $Loss_{SPE}$, here $\mu$ represent the average, $\sigma^2$ the variance or the covariance,  $c_1 =(k_1D)^2$ and $c_2 = (k_2D)^2$ are two variables that stabilise the division with a weak denominator, D is the dynamic range of the pixel values, and $k_1=0.01$ and $k_2 = 0.03$ by default.} and TV (Total Variation) to better preserve the structural similarity, the statistics and the correct amount of smoothing of the result image. Parameters $\alpha_1$, $\beta_1$ and $\gamma_1$ have been fine tuned for our task\footnote{Values of $\alpha_1$, $\beta_1$ and $\gamma_1$ are reported in our \href{https://unisannioeolab.github.io/SEN12-WATER/}{GitHub} page.}. 

\subsection{Segmentation} \label{unet_section}
The second component, used to compute the water mask, derives from the popular U-Net, commonly used for semantic segmentation tasks in computer vision. U-Net derives its name from its characteristic U-shaped architecture, which consists of a contracting path followed by an expansive path. This architecture enables the network to capture both global context and fine-grained details, making it particularly effective for tasks such as image segmentation \cite{siddique2021u}. Another notable aspect of this implementation is the inclusion of skip connections, which directly connect corresponding layers in the contracting and expansive paths. These skip connections facilitate the flow of fine-grained details from early layers to later layers, mitigating the issue of information loss commonly encountered in deep neural networks.

Our \textit{Python} implementation is based on \textit{Keras-TensorFlow} library. It has been training via a combination of the binary cross-entropy loss function, suitable for binary segmentation tasks, and the GapLoss function, specifically designed for semantic segmentation tasks, particularly in remote sensing applications. It aims to address the issue of class imbalance by focusing on the gaps between predicted and ground truth segments \cite{yuan2022gaploss}. The resulting loss is defined by:

\begin{equation}
\begin{split}
    \mathcal{L}_{SEG} &= \alpha_2\mathcal{L}_{BCE} + \beta_2 \mathcal{L}_{GAP}\\ &= -\alpha_2\frac{1}{N} \sum_{i=1}^{N} \left[y_i \log(\hat{y}_i) + (1 - y_i) \log(1 - \hat{y}_i)\right]\\ &+\beta_2 \frac{1}{N}\sum_{i=1}^{N}\left( \frac{(\hat{y}_i-y_i)^2}{(\hat{y}_i+y_i)^2 + \epsilon}\right)
\end{split}
\end{equation}

where $y_i$ is the ground truth label, $\hat{y}_i$ is the predicted probability, and $N$ is the number of pixels.  Values for $\alpha_2$ and $\beta_2$ have been fine-tuned\footnote{Values of $\alpha_2$ and $\beta_2$ are reported in our \href{https://unisannioeolab.github.io/SEN12-WATER/}{GitHub} page.}. The use of both losses allows to benefit from the robust classification capabilities of Cross-Entropy Loss and the precise boundary delineation of GapLoss. This last point is extremely important for out method, since the correct identification, at pixel-level, of water impacts on the water volume computation as detailed in the next sessions.

\begin{figure*}[!h]
    \centering
        \centering
        \includegraphics[width=\textwidth]{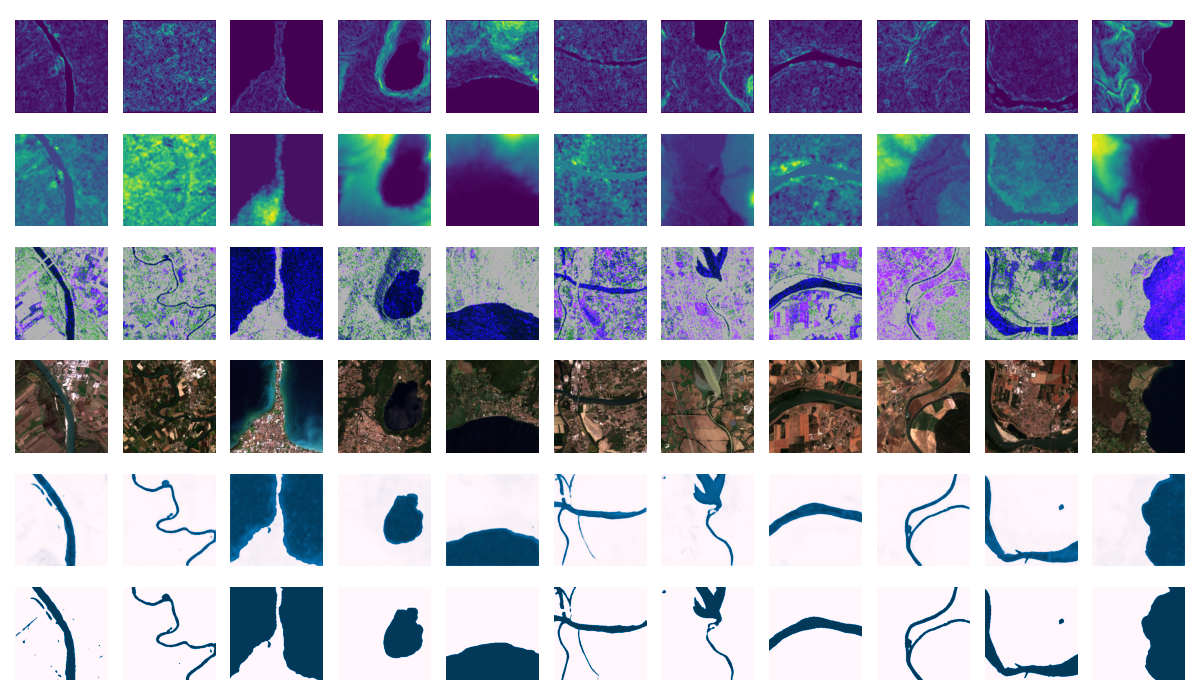}
        \label{fig:visual_comparison}
    \caption{Visual comparison of our method output (before the time-series analysis block), with respect to input data (S1 \& S2) and reference data. Each row represents, $1)$ S1 data, $2)$ S2 data, $3)$ method results, 4) ground truth.}
\end{figure*}

\subsection{Next-frame prediction task}
The last component of our method, aims at forecasting the water mask in a next-frame prediction setting. Specifically, this module forecast the water mask two months ahead using data gathered from the preceding $14$ months. These initial experiments serve as a foundational step in evaluating the introduced dataset and will serve as a basis for further exploration in future research. Three DL frameworks are investigated in our study: 1)  ConvLSTM, 2) Bidirectional ConvLSTM, and 3) a Time Distributed CNN. 

Beyond achieving accurate predictions, next-frame prediction holds practical significance. Being able to predict changes in water levels two months in advance is crucial for various applications, particularly in agriculture. For instance, anticipating water scarcity can help farmers plan their irrigation schedules and crop selection, potentially mitigating the adverse impacts of droughts on crop yields. Accurate water level predictions can also aid in water resource management, enabling authorities to make informed decisions about water allocation during critical periods.

The loss function that regulates this component is composed of three losses:

\begin{equation}
    \begin{split}
        \mathcal{L}_{NEXT}&=\alpha_3\mathcal{L}_{MSE} + \beta_3\mathcal{L}_{SSIM} + \gamma_3\mathcal{L}_{TSL}\\ &=\alpha_3\mathcal{L}_{MSE} + \beta_3\mathcal{L}_{SSIM}\\
        &+\gamma_3\frac{1}{T-1}\sum_{t=1}^{T-1}\left(\hat{y}_{t+1}-\hat{y}_t\right)^2
    \end{split}
\end{equation}

where $y_i$ is the ground truth label, $\hat{y}_i$ is the predicted probability, and $T$ is the size of the forecast horizon. Values for $\alpha_3$, $\beta_3$ and $\gamma_3$ have been fine-tuned \footnote{Values of $\alpha_3$, $\beta_3$ and $\gamma_3$ are reported in our \href{https://unisannioeolab.github.io/SEN12-WATER/}{GitHub} page.}. The combination of the MSE and SSIM helps in not only minimising the pixel-wise differences but also preserving the structural information of the frames. Moreover, to incorporate temporal information into the loss function for next-frame prediction, we included the Temporal Smoothness Loss (TSL) . This loss encourages the predicted frames to be temporally consistent with each other, and we want that miss-prediction of water basins.

\subsection{Time-series analysis}

Time-series analysis plays a pivotal role in understanding temporal patterns within datasets, offering invaluable insights into trends, fluctuations, and underlying dynamics over time. In this context, the utilisation of time-series techniques becomes particularly pertinent in elucidating the evolving nature of environmental phenomena, such as water dynamics. 

We presented the extraction of water masks from the dataset and the forecast of these, now by analysing these sequences, we aim to discern discernible trends in the fluctuating pixel count of water over time. The ambition is to discern whether variations in the volume of water stem from seasonal trends or potentially signify broader shifts attributable to climate change. This effort not only underscores the importance of robust time-series methodologies in environmental analysis but also underscores the significance of discerning nuanced patterns in water dynamics for informed decision-making and environmental management. 

Focusing on what is proposed here, the water volume variation over time is obtained by properly combining the water masks with a DEM $V_{total} = \sum (h_{i,j} \cdot w_{i,j} \cdot A)$. Where:  $h_{i,j}$ is the height of the basin at pixel $(i,j)$ extracted from the DEM, $w_{i,j}$ is the water mask value at pixel $(i,j)$ (0 if no water, 1 if water is present) and $A$ is the total area of a single pixel $(100m^2)$.

In support of the proposed methodology, the Appendix provides a separate study that validates the capability to extract water volumes from satellite data. Specifically, this additional study compares measurements obtained through satellite data with ground measurements trough a sensors network.

\section{Results}
This section presents the results of the proposed methodology on the SEN12-WATER dataset, establishing the first benchmark of the dataset.

Despite the proposed method has an end-to-end structure, it was found beneficial for the matter of scientific discussion to retrieve the performance of the main sub-modules. Regarding speckle filtering, the results are comparable with the one reported in Sebastianelli et Al. \cite{sebastianelli2022speckle} and to not overburden the paper, are not reported here, but they can be found on our \href{https://unisannioeolab.github.io/SEN12-WATER/}{GitHub} page. 

\paragraph{Results for Segmentation}
to evaluate segmentation performances, they were used standard accuracy metrics including Intersection over Union (IoU) score, Precision, and Recall \cite{del2021artificial}. Performance metrics are computed as \textit{weighted average}, where the \textit{support}, which represents the number of true instances for each label, is utilised as weight. By giving greater importance to labels with higher support, this approach addresses the issue of dataset imbalance.
\begin{table}[htbp]
    \centering
    \caption{Comparison of the proposed segmentation performances with SOTA.}
    \label{tab:res_segmentation2}
    \resizebox{1\columnwidth}{!}{

    \begin{tabular}{llll}
        \hline
        \textbf{Model} & \textbf{Precision} & \textbf{Recall} & \textbf{IoU} \\
        \hline
        \textbf{Our Model} & 0.96 & 0.97 & 0.94 \\
        %\hline
        Wieland et al. \cite{wieland2023s1s2} & 0.98 & 0.96 & 0.94 \\
        \hline
    \end{tabular}
    }
\end{table}
Results presented in Table \ref{tab:res_segmentation2} underscore the robustness of the proposed framework, especially when compared to other state-of-the-art models such as the one proposed by Wieland et al. \cite{wieland2023s1s2}, which achieved a comparable IoU of 0.94 but with slightly different data configurations.
The Precision and Recall metrics further support the effectiveness of the proposed method, particularly in scenarios where accurate water body segmentation is crucial for environmental monitoring and management.

\paragraph{Ablation study}
for the segmentation task Various combinations of inputs were tested. The findings are presented in Table \ref{tab:res_segmentation}. Specifically, Sentinel-1 data incorporating both VV and VH polarizations (after de-speckling), along with Sentinel-2 data comprising of R, G, B, and NIR bands, were employed. Furthermore, the advantages of utilizing slope, and elevation (Slo-El) were assessed. 
\begin{table}[htbp]
    \centering
    \caption{Results for the segmentation task.}
    \label{tab:res_segmentation}
    \resizebox{1\columnwidth}{!}{
    \begin{tabular}{llll}
        \hline
        \textbf{Input data} & \textbf{Precision} & \textbf{Recall} & \textbf{IoU} \\
        \hline
        S1 & 0.94 & 0.94 & 0.89 \\
        S2 & 0.96 & 0.96 & 0.93 \\
        S1+Slo+El & 0.96 & 0.96 & 0.92\\
        S2+Slo+El & 0.96 & 0.96 & 0.93 \\
        S1+S2+Slo+El & 0.96 & 0.97 & 0.94 \\
        \hline
    \end{tabular}
    }
\end{table}
The results demonstrate significant accuracy in identifying water basins. Specifically, the integration of S1 and S2 data, along with slope and elevation information, yielded the highest IoU score of 0.94.
The slight differences in Precision and Recall across different input combinations suggest that while S2 data contributes significantly to segmentation accuracy, the inclusion of SAR data from S1 is crucial for maintaining high Recall, particularly in cloudy or otherwise challenging conditions where optical data alone may be insufficient.

\begin{figure*}[!ht]
    \centering
    \resizebox{2\columnwidth}{!}{
    \begin{tabular}{cc}
         \includegraphics{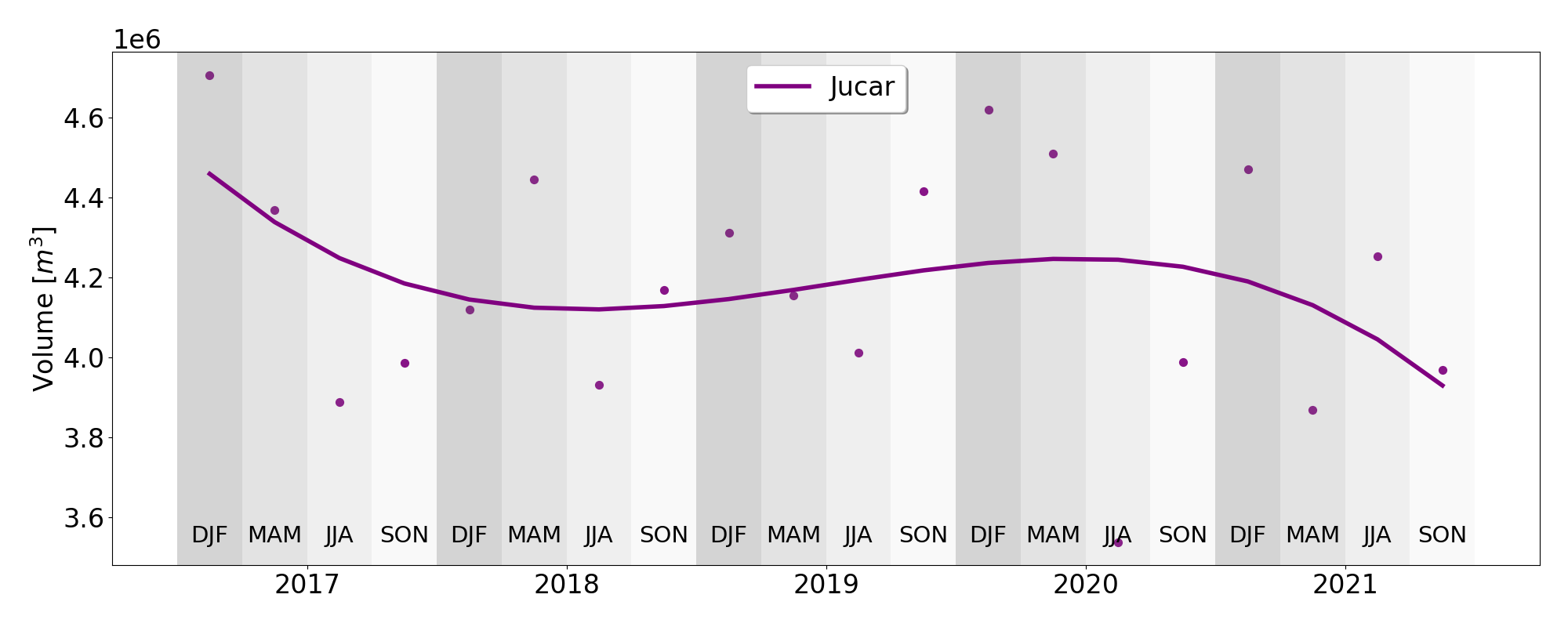} & \includegraphics{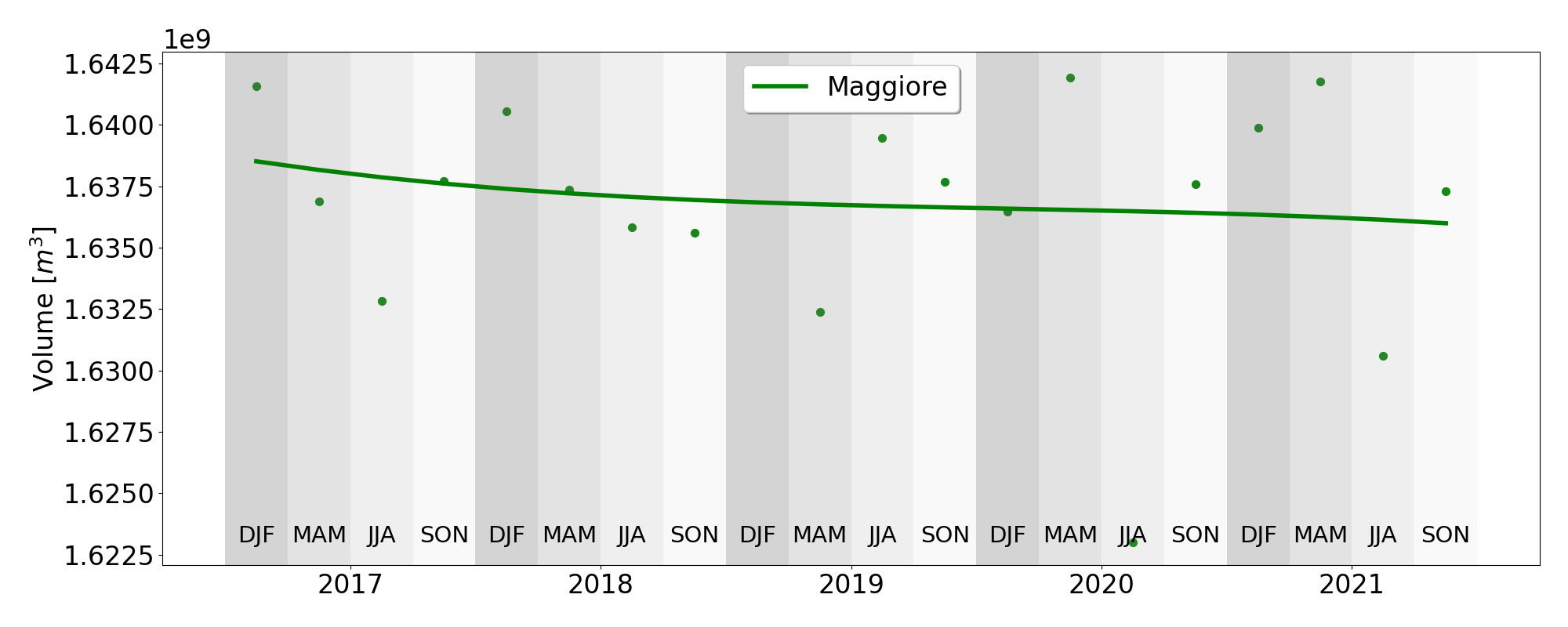} \\
         \includegraphics{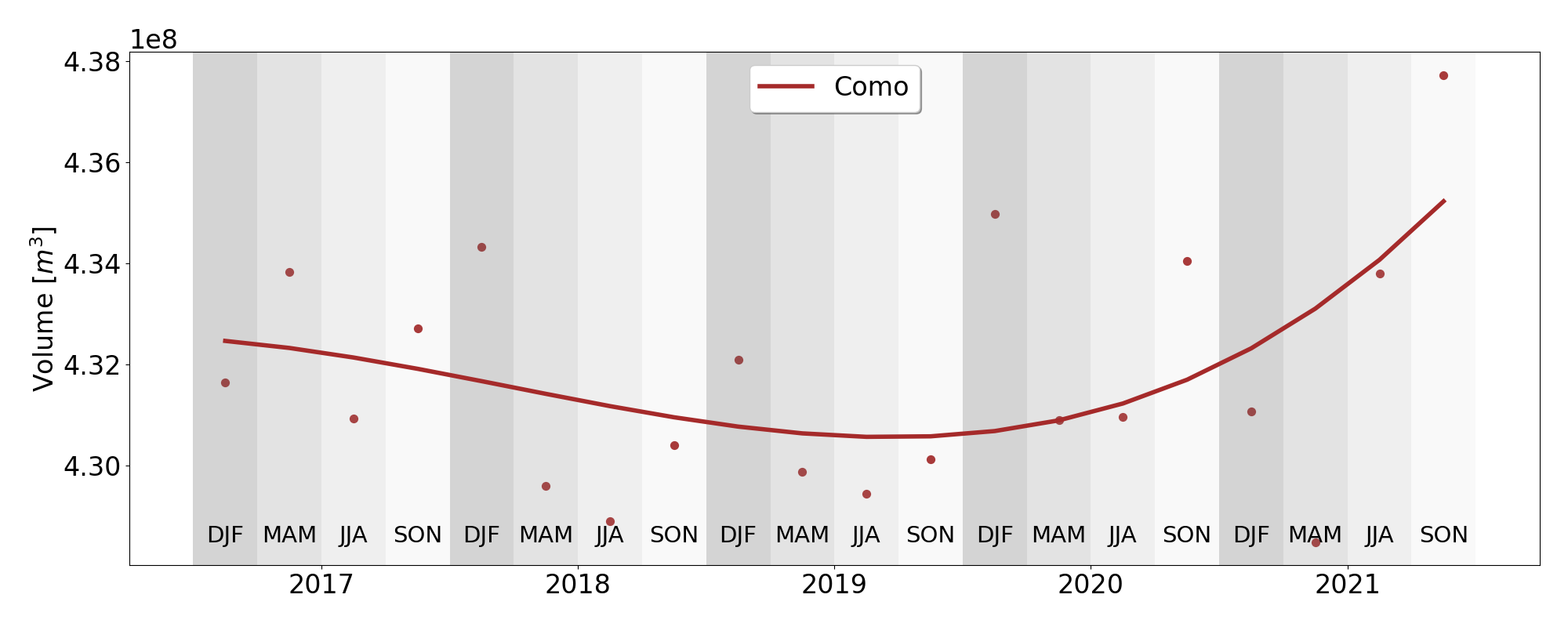} & \includegraphics{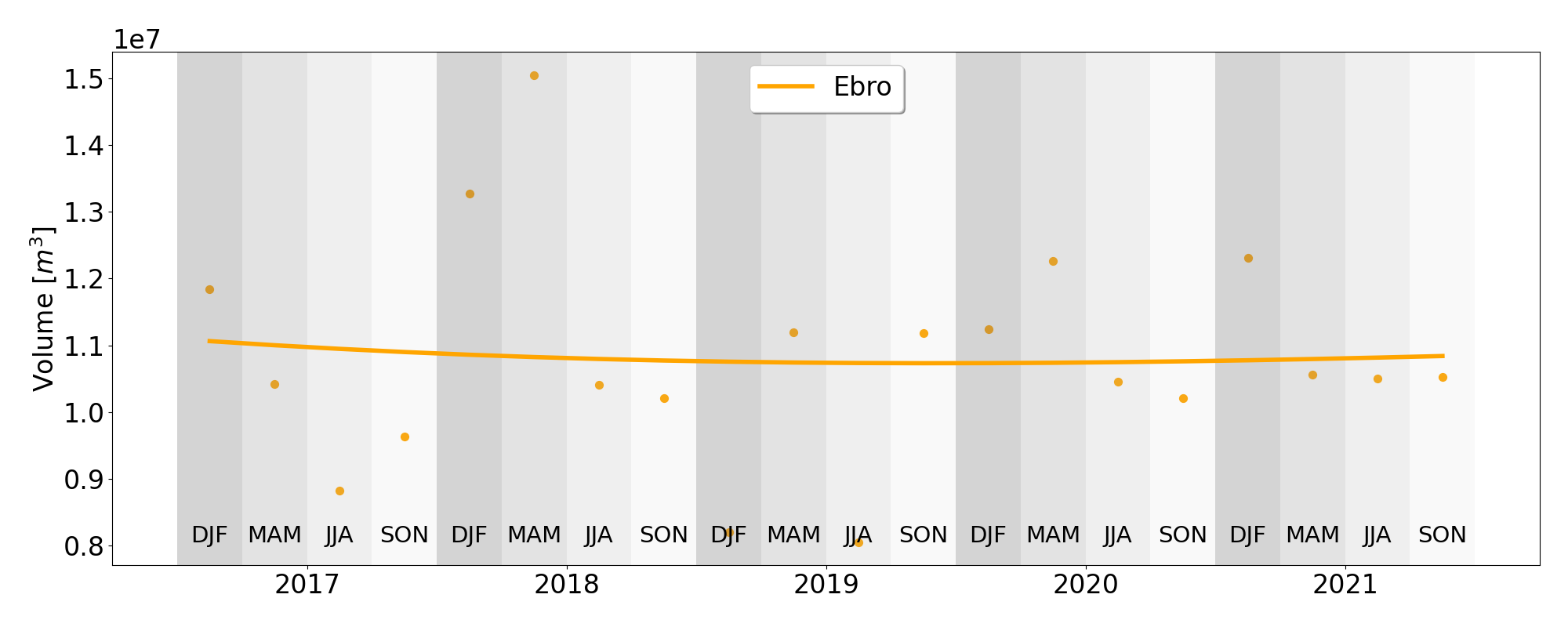} 
    \end{tabular}}
    \caption{Investigation on the temporal trends in the water volume variation.}
    \label{fig:water_curves}
\end{figure*}

\paragraph{Results for Next-Frame Prediction}
To assess the effectiveness of the proposed next-frame prediction model, they were evaluated the following metrics: Mean Squared Error (MSE),  Peak Signal-to-Noise Ratio (PSNR), Structural Similarity Index (SSIM), as already done in the initial work SEN2DWATER \cite{mauro2023sen2dwater}.
With the proposed methodology an image is foretasted based on a sequence of past acquisitions. The implemented code, available post-publication on the \href{https://unisannioeolab.github.io/SEN12-WATER/}{GitHub page} of the authors, is modular and facilitates easy testing of various networks with different parameter combinations, such as the length of the time series and the predicted water mask. The GitHub page also hosts TensorBoard records for the models utilized in this study, ensuring the reproducibility of this work. To optimize memory usage, they were provided only the final simulation results for each model in these records, encompassing model weights as well as numerical and visual outcomes obtained post-training.
The metrics obtained for the next frame prediction task have significantly improved compared to the values obtained in \cite{mauro2023sen2dwater}, and this is attributable to the more informative water mask achieved through DL-based techniques.

\begin{table}[!h]
    \centering
    \caption{Averaged numerical results on validation set through our new proposed methodology for next-frame prediction task.}\label{tab:results}
    \resizebox{1\columnwidth}{!}{
    \begin{tabular}{llll}
    \hline
    %\toprule
    \textbf{Score} & \textbf{ConvLSTM} & \textbf{Bi-ConvLSTM} & \textbf{TD-CNN} \\
    \hline
    %\midrule
    MSE $\downarrow$ & $0.013  $ & $ 0.082  $ & $ 0.007  $\\
    SSIM $\uparrow$  & $ 0.836   $ & $ 0.847  $ & $ 0.880 $\\
    PSNR $\uparrow$  & $ 62.231$ & $ 57.138 $ & $63.828 $\\
    %\bottomrule
    \hline
    \label{tab:next_frame}
    \end{tabular}}    
\end{table}
Moreover, Table \ref{tab:next_frame} underlines that Time Distributed Convolutional Neural Network (TD-CNN) outperforms other models such as ConvLSTM and Bi-ConvLSTM. The lower MSE and higher SSIM achieved by the TD-CNN indicate its superior capability in preserving spatial and temporal consistency in the predicted water masks. This is particularly important for applications such as drought prediction and water resource management, where even minor inaccuracies in water volume estimation can lead to significant implications for agricultural planning and disaster preparedness.

\paragraph{Results for time-series analysis}
the temporal trends in water volume variation, as illustrated in Figure \ref{fig:water_curves}, provide valuable insights into the impact of climate variability on water resources. Specifically, the graph presents an analysis of temporal trends in water volume variation for four different water bodies: Júcar, Maggiore, Como, and Ebro, between 2017 and 2021.
For Jucar, a slight downward trend in total water volume is observed, decreasing from around 4.6 million cubic meters in 2017 to approximately 3.8 million cubic meters in 2021.
In case of Maggiore the water volume remains relatively stable, with a slight decline from 1.64 billion cubic meters in 2017 to about 1.63 billion cubic meters in 2021.
The curve for Como initially shows a downward trend until 2020, followed by a recovery in 2021, with the volume increasing from around 4.30 to 4.35 billion cubic meters.
For Ebro, the water volume shows minimal variation, remaining stable around 1.1-1.2 million cubic meters over the analyzed period.
In summary, the trends vary across the basins: while Júcar and Como exhibit significant fluctuations, Maggiore shows only a slight decline, and Ebro remains largely unchanged.
The observed reductions in water volumes during certain periods are indicative of the increasing severity of drought conditions, which aligns with broader climate change trends. This trend analysis not only validates the utility of the SEN12-WATER dataset in capturing critical hydrological changes but also highlights its potential for use in long-term climate resilience strategies. However, further in-depth analyses are necessary before the causes of these reductions in water volumes can be deduced.

\section{Discussion}
The SEN12-WATER dataset stands out for its innovative integration of multisource and multitemporal data, combining Sentinel-1 SAR, Sentinel-2 multispectral optical data, along with slope and elevation information. 
This combination allows for effective water body monitoring under various environmental conditions, overcoming limitations such as cloud cover that affect optical data. The inclusion of both radar data, unaffected by weather, and high-resolution optical data enhances the dataset's robustness for analyzing seasonal and long-term water dynamics, which is crucial for understanding the impacts of climate change on water resources. Despite its advantages, the SEN12-WATER dataset has some limitations. It is primarily focused on Italy and Spain, which restricts its applicability to regions with different environmental and climatic conditions. Expanding the dataset to include diverse geographical areas, especially those prone to droughts or facing different water management challenges, could address this limitation. Additionally, while the dataset provides bi-monthly samples, critical changes in water dynamics may occur within shorter timescales. Increasing the sampling frequency, particularly during critical periods, would offer a more detailed understanding of these dynamics. Although the integration of SAR data helps mitigate cloud cover issues, there may still be periods with data gaps due to other operational limitations. To enhance data completeness, integrating additional satellite sources or employing synthetic data generation techniques could be beneficial.\\

Furthermore, the proposed methodology, leveraging advanced DL techniques such as U-Net for water body segmentation,  enhances the accuracy of water detection and segmentation by using a comprehensive dataset that integrates multiple data sources. Moreover, the use of a ResNet model for speckle noise removal from SAR images improves the quality of the radar data, which is crucial for accurate water body segmentation. Additionally, the combination of segmentation and prediction models (such as the TD-CNN), provides a robust framework for both current analysis and future water dynamics forecasting, making it valuable for proactive drought management and water resource planning. Additionally, the analysis of water volume variations over time represents a truly significant contribution to the scientific community and decision-makers. In fact, it serves as a valuable tool for addressing issues related to water resource management, such as drought. However, the proposed methodology, while robust, has some limitations. The models are specifically trained on the SEN12-WATER dataset, which may limit their generalizability to other datasets or regions. To address this, implementing transfer learning techniques or training on a more different set of datasets could enhance the models' adaptability to various conditions.
Additionally, the accuracy of the methodology depends heavily on the quality of the input data. Any noise, errors, or inconsistencies can adversely affect the model outputs. Therefore, improving data preprocessing—through advanced noise reduction techniques or data augmentation—could help mitigate these issues.

\section{Conclusion}
This study presents SEN12-WATER, a novel dataset and an end-to-end DL framework for the time series analysis of water bodies. The proposed dataset is a unique spatiotemporal datacube that combines radar polarization, elevation, slope, and multispectral optical bands, providing comprehensive data for the assessment of water dynamics. The methodology integrates advanced models, including a ResNet for speckle noise removal, a U-Net for water body segmentation, and a TD-CNN for future water mask prediction. 
The findings contribute significantly to climate change resilience and sustainable water resource management by providing actionable insights for policymakers and stakeholders. The detailed analysis of water dynamics and the ability to predict future water conditions support informed decision-making and proactive management of water resources.

%Future work will focus on further enhancing the dataset and the DL framework by incorporating additional data sources and improving model accuracy. It is also possible to explore the application of the proposed methodology to other environmental monitoring tasks, thereby broadening its impact and utility.

%In conclusion, the SEN12-WATER dataset and the associated DL framework represent a significant advancement in the field of water monitoring and assessment, offering new tools and methodologies that address current limitations and open up new possibilities for research and practical applications.

In conclusion, the results presented in this study not only demonstrate the effectiveness of the proposed framework in water body segmentation and prediction but also underscore its broader applicability in environmental monitoring. The ability to accurately predict water availability two months in advance represents a significant advancement in the field, offering critical support to policymakers and stakeholders in managing water resources more effectively. This underscores the utility of this task, demonstrating its relevance and practical application in areas such as agriculture and resource management, where early awareness of water level changes can significantly influence planning and decision-making.

\section{Appendix A}
\begin{figure*}[ht!]
    \centering
    \begin{subfigure}[b]{0.3\textwidth}
        \centering
        \includegraphics[width=\textwidth]{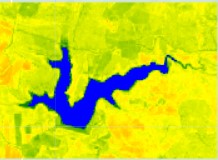}
        \caption*{(a)}
    \end{subfigure}
    \hfill
    \begin{subfigure}[b]{0.3\textwidth}
        \centering
        \includegraphics[width=\textwidth]{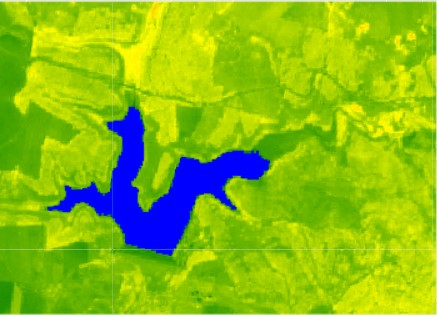}
        \caption*{(b)}
    \end{subfigure}
    \hfill
    \begin{subfigure}[b]{0.3\textwidth}
        \centering
        \includegraphics[width=\textwidth]{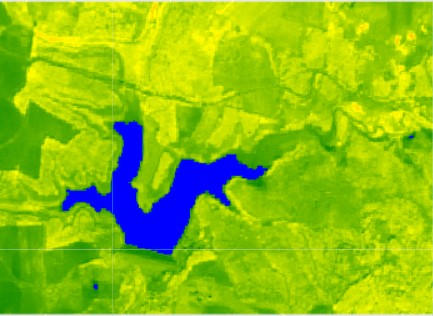}
        \caption*{(c)}
    \end{subfigure}
    \caption{Olivo Dam in different seasons: (a) spring, April 13, 2016; (b) summer, August 21, 2016; (c) autumn, October 10, 2016.}
    \label{fig:Olivo_Dam}
\end{figure*}

This appendix presents an additional study that validates the retrieval on water volume presented in the main body of the article. Specifically, this study focuses on the Olivo Dam over the time period that spans from 2016 to 2018. An example of the satellite acquisition on the Olivo dam is reported in Fig. \ref{fig:Olivo_Dam}. The surface of the water, from S2 data, after segmenting the water mask as described in Section \ref{unet_section}, has been computed via $A_{total} = \sum (w_{i,j} \cdot A)$.  Where, $w_{i,j}$ is the water mask value at pixel $(i,j)$ (0 if no water, 1 if water is present) and $A$ is the total area of a single pixel $(100m^2)$.

The theoretical data of reservoir volumes and associated areas were derived from a morphological analysis of the basin. The complete morphometric information of a hydrographic basin requires an analysis of the distribution of elementary areas that make up the basin in relation to the progression of contour lines that delimit these areas. This involves relating area information with the altimetric characteristics of the basin. The altimetric information is essential for determining various parameters that influence the kinematic characteristics of the hydrographic network, contributing to the accuracy of hydrological modeling and water resource management.

Table \ref{error_summary} presents the values extracted for the period of analysis. The first column shows the dates considered, the second column shows the basin area in square meters obtained from theoretical data, the third column shows the basin area calculated through Sentinel-2, and the fourth column shows the error.

\begin{table}[ht!]
\caption{Summary of comparison analysis between theoretical (second column) and satellite (third column) data.}
\centering
\begin{tabular}{lccc}
\toprule
\textbf{Date} & \textbf{GT [$m^2$]} & \textbf{Measured Area [$m^2$]}& \textbf{Difference}\\ 
\midrule
04/13/2016 & 512.700 & 445.214 & 67.486 \\ 
07/21/2016 & 398.100 & 352.533 & 45.567 \\ 
10/10/2016 & 361.180 & 303.055 & 58.124 \\ 
03/19/2017 & 397.320 & 342.156 & 55.163 \\ 
07/07/2017 & 354.940 & 295.035 & 59.905 \\ 
09/15/2017 & 303.440 & 260.546 & 42.894 \\ 
01/18/2018 & 377.300 & 329.973 & 47.327 \\ 
03/29/2018 & 409.125 & 345.726 & 63.398 \\ 
07/07/2018 & 401.480 & 344.057 & 57.422 \\ 
09/05/2018 & 393.680 & 335.012 & 58.668 \\ 
\bottomrule
\label{error_summary}
\end{tabular}
\end{table}

The experimental analysis of the Olivo Dam using Sentinel-2 data proves that the capability of the methodology for remote monitoring of water bodies, providing valuable data for further research and management.

\bibliographystyle{IEEEtran.bst}
\bibliography{main}

\begin{IEEEbiography}[{\includegraphics[width=1in,height=1.25in,clip,keepaspectratio]{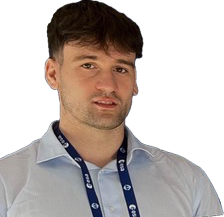}}]{Luigi Russo} Student Member, IEEE, earned his master’s degree (cum laude) in Electronic Engineering for Automation and Telecommunications from the University of Sannio, Benevento, Italy, in 2023. He is currently pursuing a Ph.D. at the University of Pavia in collaboration with the Italian Space Agency (ASI) in Rome. He has coauthored and presented papers at several prestigious conferences and received the Best Poster Award in Urban and Data Analysis as a young scientist at the 2024 European Space Agency (ESA)-Dragon Symposium. His research focuses on remote sensing and artificial intelligence for the automatic classification and analysis of satellite data, with particular emphasis on creating maps of building exposure and vulnerability to natural hazards and extreme events.
\end{IEEEbiography}
~
\begin{IEEEbiography}[{\includegraphics[width=1in,height=1.25in,clip,keepaspectratio]{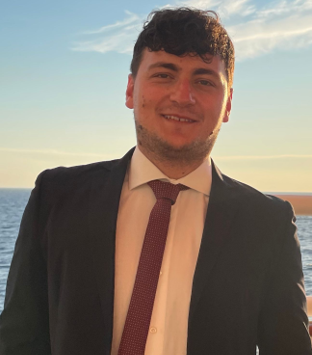}}]{Francesco Mauro} Student Member, IEEE received the master degree (with laude and career mention) in Electronic Engineering for Automation and Telecommunications in 2022 from the University of Sannio, Benevento, Italy, where he is currently working toward the Ph.D. degree. He is a Visiting Researcher with the $\Phi$-lab, European Space Agency, Frascati, Italy, and collaborates with the $\Phi$-lab on topics related to Quantum Machine Learning applied to Earth observation. He has coauthored several papers for the sector of remote sensing and has presented his workd in reputed international conferences. His research interests include remote sensing and satellite data analysis, Artificial Intelligence and Quantum Machine Learning techniques for Earth observation. He is also member of the GRSS AdCom, serving as Chapter Coordinator for Region 8.
\end{IEEEbiography}\vspace{-2\baselineskip}
~
\begin{IEEEbiography}[{\includegraphics[width=1in,height=1.25in,clip,keepaspectratio]{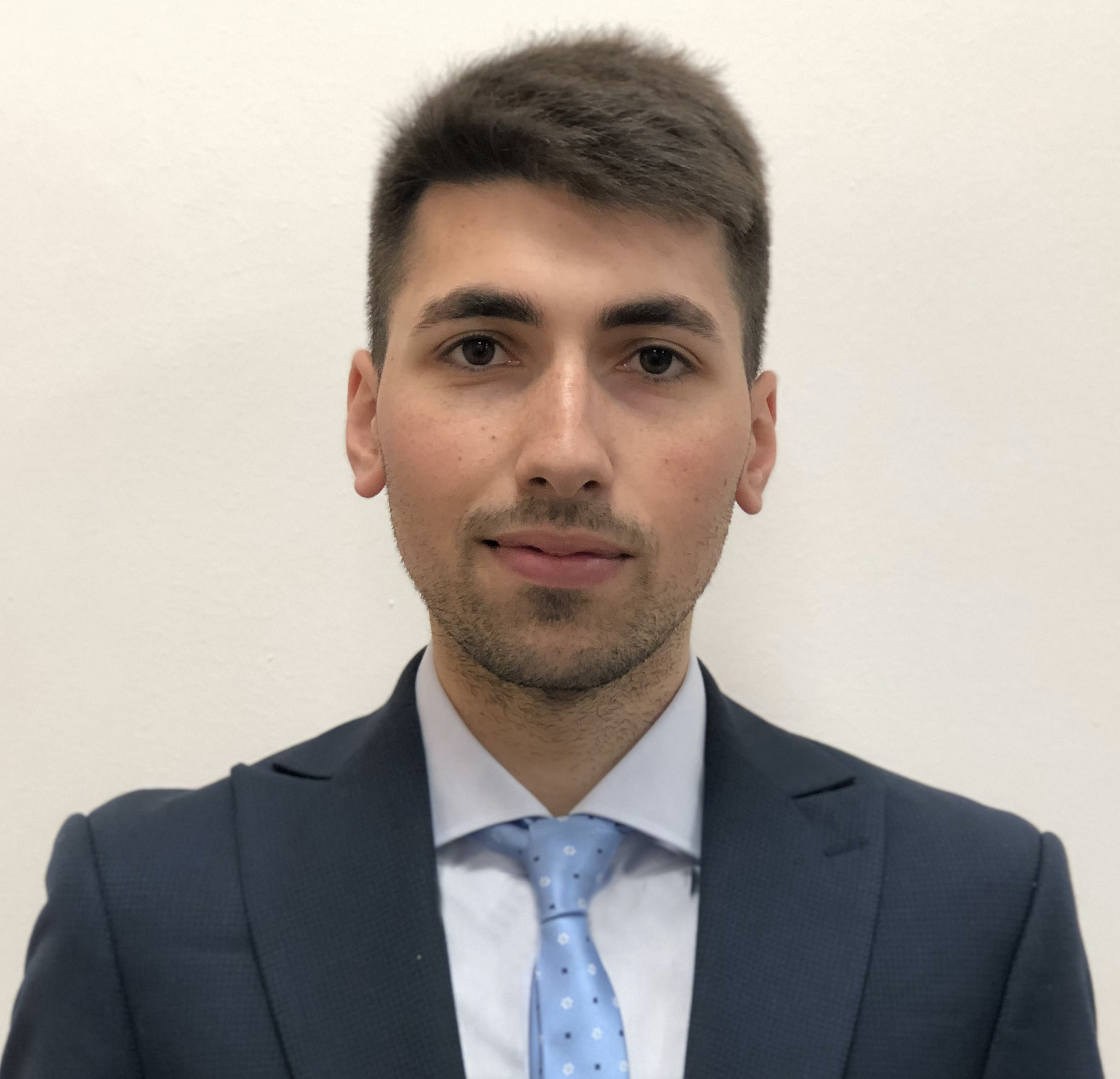}}]{\href{https://alessandrosebastianelli.github.io/}{Alessandro Sebastianelli}} received the degree (cum laude) in Electronic Engineering for Automation and Telecommunications at the University of Sannio, Benevento, Italy, in 2019, where he also pursued the Ph.D. degree. His research topics mainly focus on remote sensing and satellite data analysis, artificial intelligence (AI) techniques for Earth observation, data fusion and quantum machine learning. He has coauthored several articles to reputed journals and conferences for the sector of remote sensing. Ha has been a Visiting Researcher with $\Phi$-lab, European Space Agency ESA/European Space Research Institute ESRIN, in Frascati, Italy, and still collaborates with. He has won an ESA OSIP proposal in August 2020. He received an IEEE award for one of the best three theses in Geoscience and Remote Sensing (GRS) in Italy. Currently, he works as a Research Fellow in Quantum Computing for Earth Observation at the  $\Phi$-lab, ESA. 
\end{IEEEbiography}\vspace{-2\baselineskip}
~
\begin{IEEEbiography}[{\includegraphics[width=1in,height=1.15in,clip,keepaspectratio]{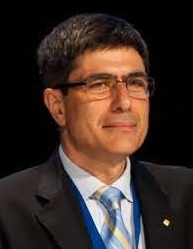}}]{Paolo Gamba}
IEEE Fellow, received the Laurea (cum laude) and Ph.D. degrees in electronic engineering from the University of Pavia, Pavia, Italy, in 1989 and 1993, respectively.
He is a Professor of telecommunications with the University of Pavia, where he leads the Telecommunications and Remote Sensing Laboratory and serves as a Deputy Coordinator of the Ph.D. School in Electronics and Computer Science. He has been invited to give keynote lectures and tutorials in several occasions about urban remote sensing, data fusion, EO data, and risk management.
Dr. Gamba has served as the Chair for the Data Fusion Committee of the IEEE Geoscience and Remote Sensing Society from 2005 to 2009. He has been elected in the GRSS AdCom since 2014. He was also the GRSS President. He had been the Organizer and Technical Chair of the biennial GRSS/ISPRS Joint Workshops on Remote Sensing and Data Fusion over Urban Areas from 2001 to 2015. He has also served as the Technical Co-Chair of the 2010, 2015, and 2020 IGARSS Conferences, Honolulu, HI, USA, and Milan, Italy, respectively. He was the Editor-in-Chief of the IEEE Geoscience and Remote Sensing Letters  from 2009 to 2013.
\end{IEEEbiography}\vspace{-2\baselineskip}
~
\begin{IEEEbiography}[{\includegraphics[width=1in,height=1.15in,clip,keepaspectratio]{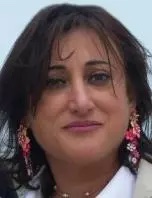}}]{Silvia Liberata Ullo} IEEE Senior Member, President of IEEE AES Italy Chapter, Industry Liaison for IEEE Joint ComSoc/VTS Italy Chapter since 2018, National Referent for FIDAPA BPW Italy Science and Technology Task Force (2019-2021). Member of the Image Analysis and Data Fusion Technical Committee (IADF TC) of the IEEE Geoscience and Remote Sensing Society (GRSS) since 2020. Graduated with laude in 1989 in Electronic Engineering at the University of Naples (Italy), pursued the M.Sc. in Management at MIT (Massachusetts Institute of Technology, USA) in 1992. Researcher and teacher since 2004 at University of Sannio, Benevento (Italy). Member of Academic Senate and PhD Professors’ Board. Courses: Signal theory and elaboration, Telecommunication networks (Bachelor program); Earth monitoring and mission analysis Lab (Master program), Optical and radar remote sensing (Ph.D. program).  Authored 90+ research papers, co-authored many book chapters and served as editor of two books. Associate Editor of relevant journals (IEEE JSTARS, MDPI Remote Sensing, IET Image Processing, Springer Arabian Journal of Geosciences and Recent Advances in Computer Science and Communications). Guest Editor of many special issues. Research interests: signal processing, radar systems, sensor networks, smart grids, remote sensing, satellite data analysis, machine learning and quantum ML.
\end{IEEEbiography}

\end{document}